\documentstyle[epsfig,psfig,epsf]{elsart}
\begin{document}
\begin{frontmatter}
\title{A System with Multiple Liquid-Liquid\\ Critical Points}

\author{Sergey V. Buldyrev}\footnote{buldyrev@bu.edu}\ and
\author{H. Eugene Stanley}

\address{Center for Polymer Studies and Department of Physics\\
Boston University, Boston, MA 02215, USA}

\date{bs.tex ~~~ 4 April 2003}

\begin{abstract}

We study a three-dimensional system of particles interacting via
spherically-symmetric pair potentials consisting of several
discontinuous steps. We show that at certain values of the parameters
desribing the potential, the system has three first-order phase
transitions between fluids of different densities ending in three
critical points.

\end{abstract}

\end{frontmatter}

Much attention has been focused on the topic of liquid-liquid phase
transitions \cite{volga}. Recently a liquid-liquid phase transition was
found experimentally in phosphorus \cite{Katayama}. Liquid-liquid phase
transitions also can exist in other materials with tetrahedral symmetry
such as carbon \cite{Togaya,Glossili}, silica, \cite{Saika} and silicon
\cite{Borick,Angell}.

The possibility of a sharp amorphous-amorphous transition between low
density amorphous ice (LDA) and high density amorphous ice (HDA) has
been known for a long time \cite{Mishima}. Computer simulations of the
ST2 water model \cite{Poole} suggest that the transition line between
LDA and HDA can be extended to higher temperatures above the glass
transition, where it would become a first order liquid-liquid phase
transition line ending at a critical point at about 200K
\cite{Mishima1}. However, in this temperature range, water immediately
crystallizes and this phase transition cannot be observed in direct
experiments.  More recently, a distinct amorphous phase of water, called
{\it very high-density amorphous ice}, has been observed experimentally
\cite{Loerting,Finney,Klug}. New simulations of the ST2 model
\cite{Geiger} in the restricted NPT ensemble suggest that instead of one
previously observed liquid-liquid phase transition, there might be three
liquid-liquid phase transition lines between four liquids with different
densities roughly corresponding to the densities of the three known
amorphous ices.  It has been suggested \cite{Brazhkin} that systems with
rich polymorphic solid phase diagrams may have several liquid phases
with local particle arrangements corresponding to various crystalline
structures.

We propose here a simple model with at least two liquid-liquid
phase transitions in addition to the known liquid-gas phase transition.
This model is based on the double step soft-core potential model 
previously used to study systems with two fluid-fluid critical points
\cite{Franzese,Franzese1,Buldyrev,Skibinsky}. Two fluid-fluid critical
points were also observed by Monte Carlo simulations of a similar
model with a continuous soft-core potential \cite{Jagla}. 

We perform constant volume $V$ collision-driven 
molecular dynamic simulations \cite{Franzese1,Buldyrev,Rapaport}
on a system of $N=1728$ spherically 
symmetric particles interacting via the discontinuous pair potential
\begin{equation}
U(r)=\left\{
\begin{array}{lll}
\infty & \mbox{  for  } & r < a \\
U_1    & \mbox{  for  } & a\leq r < b_1 \\
U_2    & \mbox{ for  } & b_1\leq r <b_2  \\
-U_A    & \mbox{ for  } & b_2\leq r <c  \\
0      & \mbox{  for  } & c\leq r ~~~, 
\end{array}\right.
\label{Uofr}
\end{equation}
where $U(r)$ is a potential energy of a pair of particles at distance
$r$, $U_A>0$ is the energy of attraction, $U_1=3U_A$, $U_2=U_A$,
$b_1=1.4a$, $b_2=2.1a$, and $c=2.8a$. (see Figure 1, inset).  We keep
temperature $T$ constant using a modified Berendsen method and compute
pressure $P$ the same way as in \cite{Franzese,Franzese1}, averaging it over
$2\cdot 10^4~(ma^2/U_A)^{1/2}$ time units, where $m$ is particle
mass. Throughout this paper, we measure density, pressure, and
temperature in units of $a^{-3}$, $U_Aa^{-3}$, and $U_A/k_B$,
respectively, where $k_B$ is the Boltzmann constant. We investigate
several state points for systems with different densities $\rho\equiv N/V$
along several isotherms (see Fig.~1) and find three distinct sets of
van der Waals loops, each ending in a specific critical point
characterized by an isotherm with an inflection point. 

\begin{figure}
\includegraphics[width=10cm,height=13cm,angle=270]{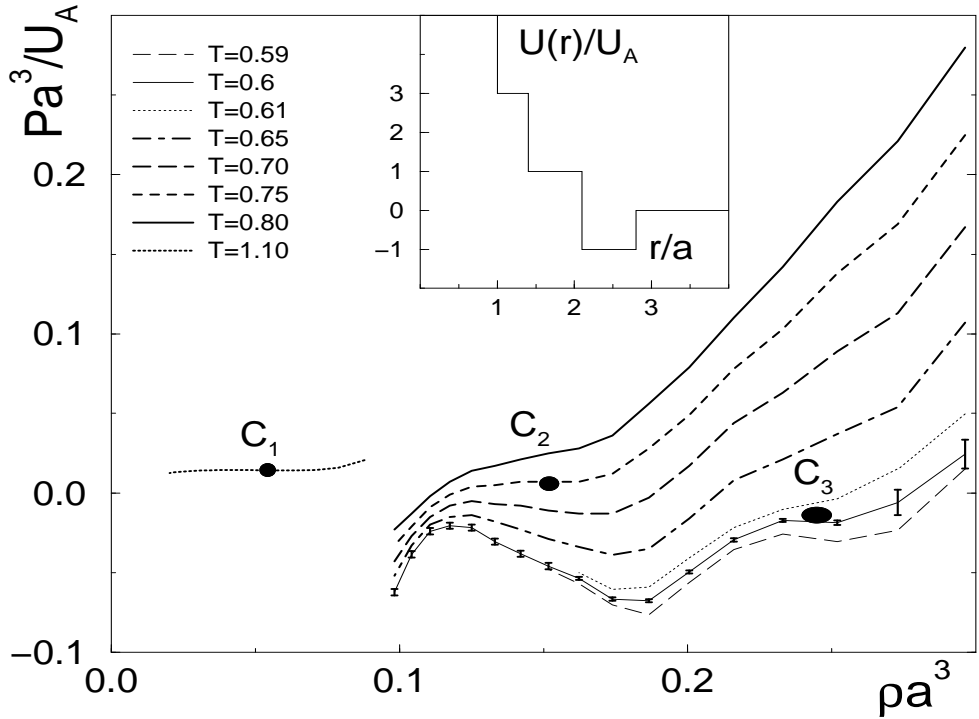}
\caption{Isotherms and critical points for the system of spherically 
symmetric particles interacting via the potential shown as an inset.
\label{f.1}}
\end{figure}

Note that the isotherm for $T=0.60$ almost entirely corresponds to a
stretched metastable liquid with $P<0$. In addition, at this
temperature, the system spontaneously crystallizes for
$\rho>0.23$. Nevertheless, we are able to find equilibrium liquid
pressures by averaging pressure before the crystallization onset, using
the same technique as in \cite{Franzese1}. We quench five independent
configurations equilibrated at T=0.9 to T=0.6, skip initial 800 time
units (the correlation time at this temperature), and average the
pressure until the onset of crystallization, indicated by the sharp
decrease of potential energy. The average existence time $\langle t
\rangle$ of the metastable liquid is $\langle t \rangle=1.5\cdot 10^4$
time units at $\rho=0.23$ , $\langle t \rangle=4.5\cdot 10^3$ at
$\rho=0.25$, $\langle t \rangle=1.5\cdot 10^3$ at $\rho=0.27$, and
$\langle t \rangle = 10^3$ at $\rho=0.30$. The error bars of the
pressure are assumed to be equal to the standard deviation of the
pressure measurements for five independent runs.

Accordingly, we find a liquid-gas critical point $C_1$ at $T_1=1.10$,
$P_1=0.014$, $\rho_1=0.054$, a stable high density critical point $C_2$
at $T_2=0.75$, $P_2=0.0056$, $\rho_2=0.152$, and a third very high
density critical point $C_3$ at $T_3=0.60$, $P_3=-0.014$,
$\rho_3=0.245$. The third critical point is metastable with respect to
gas-crystal coexistence, and by varying the parameters of the potential
\cite{Skibinsky}, one should be able to shift the position of the third
critical point to a stable region of the phase diagram.

\begin{figure}
\includegraphics[width=10cm,height=13cm,angle=270]{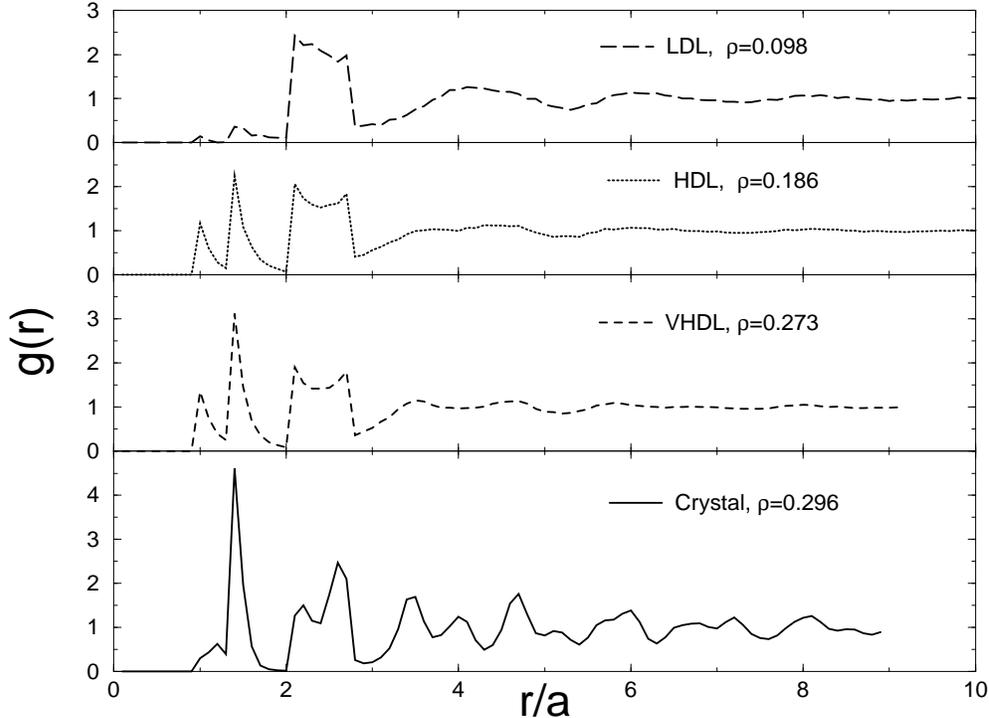}
\caption{Radial distribution functions for a system shown in Fig.~1 at
temperature $T=0.59$ at four different densities, corresponding to
three different liquids and a crystal.
\label{f.2}}
\end{figure}

Figure 2 shows radial distribution functions for the three liquids at
temperature $T=0.59$, corresponding to the low density liquid (LDL)
with density 0.098$a^{-3}$, high density liquid (HDL) with density
0.186$a^{-3}$, very high density liquid (VHDL) with density
0.272$a^{-3}$ and a crystal with density 0.296$a^{-3}$. 
One can see a dramatic difference between the structures
of LDL and HDL.  

Integrating radial distribution function for LDL, we find that
11 neighbors stay in the attractive well, almost not
penetrating into soft cores.

In contrast, in HDL, particles penetrate into a larger soft core. 
The average number of neighbors in the smaller soft core is 0.6. The
average number of neighbors in the larger soft core is 2.7. The number
of particles in the attractive well also increases to 17.  

The difference between the structures of HDL and VHDL is less
pronounced. Nevertheless, one can see the development of an extra peak
for VHDL at a distance 3.5$a$. In addition, the number of particles in the
smaller and larger soft cores increases to 1 and 5, respectively, and
the number of particles in the attractive well increases to 23.

The choice of parameters is motivated by our preliminary studies of
simpler systems with $U_1=\infty$, which have a gas-liquid critical
point $\tilde C_1$ ( $T=1.06,~P=0.013,~\rho=0.052$) and a stable
liquid-liquid critical point $\tilde C_2$ ($T=0.67,~P=
0.059,~\rho=0.144$ ) (see Fig. 3). The potential of this system is the
soft-core potential \cite{Franzese}, with $b_1$ playing the role of the
hard core and $b_2$ playing the role of the soft core. Note that the
positions of these critical points are very similar to those of $C_1$
and $C_2$ in the three-step potential. Thus, adding a second soft core
$a< b_1 < b_2$ to the potential, creates a third critical point at
higher densities without a significant change in the position of the two
original critical points.

\begin{figure}
\includegraphics[width=10cm,height=13cm,angle=270]{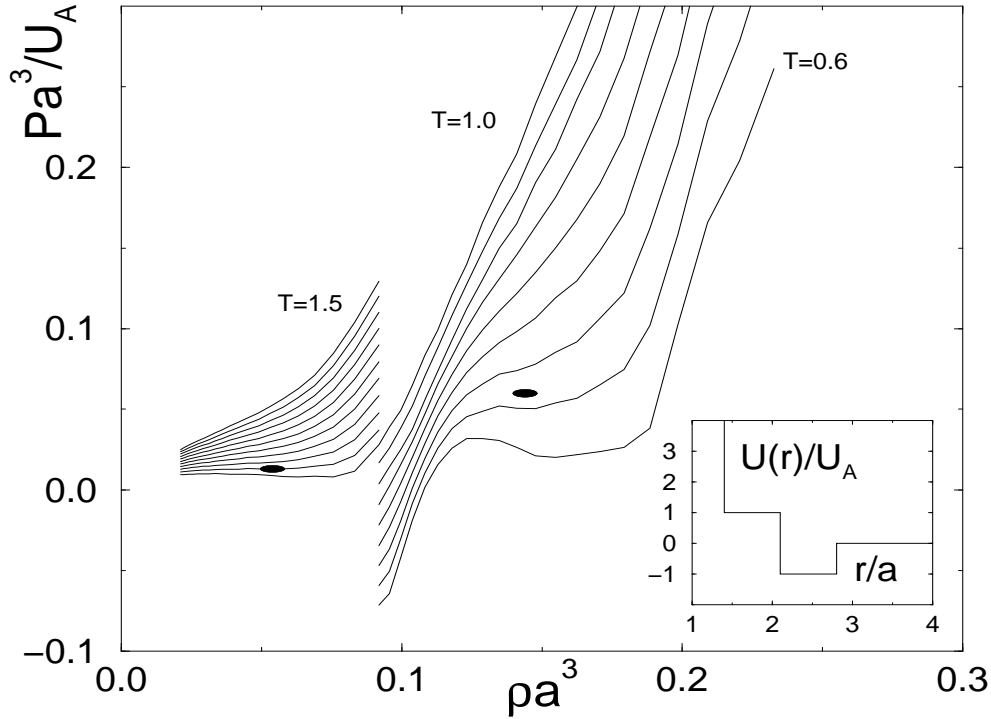}
\caption{Isotherms and critical points for temperatures $T=$1.5, 1.45,
1.4, 1.35, 1.30, 1.25, 1.20, 1.15, 1.10, 1.05, 1.00, 0.95, 0.90, 0.85,
0.80, 0.75, 0.70, 0.65, 0.60 for a simplified potential shown in the
inset with one soft core $b_2$ and a hard core $b_1=1.4a$.  Isotherms
for $T>1.00$ are computed only for small densities, while isotherms for
$T<1.00$ are computed only for large densities.  The isotherm
corresponding to $T=1.00$ is computed for the entire range of
densities. The systems for $\rho>0.23$ spontaneously crystallize at
$T<0.75$ into crystalline structures similar to to the crystalline
structure of the system with two soft cores shown in Fig. 2.
\label{f.3}}
\end{figure}

One can speculate that adding more steps to a potential can create more
critical points at different densities corresponding to the
interparticle distances of these steps if the parameters of these steps
are carefully selected.

In summary, we show that systems interacting via spherically symmetric
potentials with three characteristic repulsive distances playing the
role of soft cores may have three liquid phases of three increasing
densities characterized by the penetration of particles into a soft core
of smaller diameter.

\subsubsection*{Acknowledgments}

We thank V. V. Brazhkin, A. Geiger, G. Franzese, T. Loerting, and
A. Skibinsky for helpful discussions and the NSF Chemistry Division
(CHE-0096892) for support.

\end{document}